\documentclass[12pt]{article}
\usepackage{epsfig}
\begin{document}
\noindent {\Large \bf Reverse Brazil Nut Problem: Competition \\
between Percolation and Condensation}
\vskip 1.0 true cm
\centerline{ \bf Daniel C. Hong and Paul V. Quinn }
\vskip 0.2 true cm
\centerline{ Physics, Lewis Laboratory,
Lehigh University, Bethlehem, Pennsylvania 18015}
\vskip 0.3 true cm
\centerline{ \bf Stefan Luding}
\vskip 0.2 true cm
\centerline{ Institut f\"ur Computeranwendungen 1 Pfaffenwaldring 27, 70569
Stuttgart, Germany}
\vskip 0.3 true cm

\begin{abstract}
In the Brazil nut problem (BNP), hard spheres with larger diameters rise to
the top.  There are various explanations (percolation, reorganization,
convection), but a broad understanding or control of this effect is by no 
means achieved.  A theory is presented for
the crossover from BNP to the reverse Brazil nut problem 
(RBNP) based on a competition between the percolation effect and the
condensation of hard spheres.  The crossover condition is
determined, and theoretical predictions are compared to 
Molecular Dynamics simulations in two and three dimensions.
\end{abstract}

Rosato et al.\ [1] demonstrated via Molecular Dynamics (MD) simulations 
that hard spheres with large diameters segregate to the top
when subjected to vibrations or shaking.  In the literature,
this phenomenon is called the Brazil nut problem (BNP) [2]. Besides a 
broad experience in the applied and engineering sciences [2,3], there exist
more recent approaches to understanding this effect through model
experiments [4-7].
The BNP has been attributed to the following phenomena: the {\em percolation}
effect, where the smaller ones pass through the holes created by the 
larger ones [2], geometrical {\em reorganization}, through which small 
particles readily fill small openings below the large particles [3,5] and
global convection which brings the large particles up but does not
allow for re-entry in the downstream [4].
Since most experiments were carried out with a {\it single} large grain 
in a sea of smaller ones [4-6], it is not quite clear which
of these observed mechanisms apply for the segregation of binary mixtures.
For example, while the convection is responsible for
the rise of the {\it single} large grain, MD simulations [8] and 
hydrodynamic models [9] clearly indicate that convection cells are absent in
the bulk, but confined near the wall when the width of the container is much 
larger than the height.
Recently, Shinbrot and Muzzio [10] observed a reverse buoyancy in
shaken granular beds, where again a {\it single}
large grain in a sea of smaller grains could segregate to the {\it bottom} 
if the bed was deep and the amplitude of vibration was large.
While all these scenarios are interesting, we recognize that the basic
control parameters for each regime have not been clearly identified.
Numerical simulations of binary mixtures with external temperature 
gradients and varying ratios of gravitational acceleration and agitation 
intensity [11] show that segregation can be tuned, avoided and even inversed
under special, but non-practical boundary conditions.
However, these simulations did not examine a variation in 
mass-ratio or size-ratio and were not compared to the kinetic 
theory of binary mixtures [12].

The purpose of this Letter is to propose a new {\it condensation} driven 
segregation of binary hard spheres under gravity, 
which is conceptually 
different from the previously known convection driven [4], percolation
driven [2,3], reorganization controlled [5], entropy driven (in
the absence of gravity) [7], or inertia driven [10] 
segregation processes.
We will identify the control parameter(s) %lui?
and determine the crossover condition from the 
normal BNP to the reverse Brazil nut problem (RBNP).

The starting point 
is the observation made in an earlier paper [13]
and recently confirmed by MD simulations [14] that there exists a
critical temperature, $T_c$, below which a {\it monodisperse} 
system of hard spheres undergoes a condensation transition in 
the presence of gravity. The density profile was obtained using
the global equation of state of the system [15]. Consider, for
example, a system of elastic hard spheres with mass $m$ and 
diameter $d$ in a container.  
Let the initial layer thickness (filling height at rest when
$T=0$) be $\mu$ measured in units of $d$.
If the system is in contact with a thermal reservoir at temperature $T$,
one can estimate the dimensionless
thickness of the fluidized layer, $\Delta h$, in D dimension
by equating the kinetic energy of one grain to the potential energy
equivalent, 
$m \langle v^2 \rangle /2 = DT/2 \approx mgd\Delta h$, which yields
$\Delta h \approx DT/(2mgd)$.
One may estimate the point, $T_c$, at which
the system is fully fluidized, by setting $\Delta h \approx \mu$.  
We obtain:
$$ T_c=mgd\mu/\mu_o ~,\eqno (1)$$
where $\mu_o$ is a constant that depends on the spatial dimension 
and the underlying packing structure [16].  
$\mu_o$ was determined in Ref.\ [13] with the use of the Enskog pressure, 
while  the numerically determined values for $\mu_o$ were provided in 
Ref.\ [14].  

When $T>T_c$, the system is fully fluidized and 
the dimensionless thickness of the solid, defined as $h_{\rm solid}(T)=
\mu-\Delta h(T)$, vanishes.
On the other hand, for $T<T_c$, one finds $h_{\rm solid} \ne 0$ because
a fraction of particles condenses at the bottom. %lui forming a solid.  
Here, the solid phase refers to a hard sphere state where
each particle fluctuates around a fixed point, but is
confined in a cage so that it can not exchange its center of mass 
position with neighboring particles [14].  

Consider now a {\it binary mixture} of hard spheres with species 
A and B having mass $m_A$ and $m_B$ and diameters $d_A$ and $d_B$,
respectively, and the initial layer thicknesses are $\mu_A$ and $\mu_B$.  
From Eq.\ (1), we 
find the ratio of the critical temperatures to be
$$ \frac{T_c(A)}{T_c(B)} = \frac{m_Ad_A}{m_Bd_B}~, \eqno (2)$$
where we have assumed that $\mu_A=\mu_B$.  
Suppose the system is quenched at a temperature $T$ 
between the two critical  temperatures $T_c(A)$ and $T_c(B)$, i.e. 
$T_c(B) < T < T_c(A)$.  Then,
the hard spheres of type B are above the condensation temperature, while
the hard spheres of type A are below it. Thus at temperature $T$, particles 
of type A will try to condense,
while particles of type B remain fluidized.  Hence, type A particles 
will tend to {\it segregate} to the bottom first.   
The underlying assumption of this scenario is that particles
of type A only interact with themselves while seeing particles of type B as 
phantom particles, and vice versa.  Even though this assumption is a 
crude one, we nevertheless find that the prediction based on this assumption 
appears to work well for the segregation of binary mixtures under 
the influence of gravity.

Since the critical temperature depends on $m$ and $d$, an inversion
of the segregation should be achieved by simply altering the values 
of $m$ and $d$.  
We first present Event-Driven Molecular Dynamics data for the normal
Brazil nut problem(BNP), where particles with a larger diameter rise to the
surface.  Subsequently, we demonstrate how to {\it reverse} this phenomenon 
by quenching the system between the two condensation temperatures given by 
Eq.\ (1) and used in Eq.\ (2).

The total number of particles used in our
simulations varied from 400 to 2,000 in 2D, and 2000 to 3,600 in 3D, with 
the width of the container varying from 15 to 50 particle diameters in 2D, 
and staying fixed at 15x15 in 3D.
We refer the readers to references [17] 
for details of the algorithm, which takes into account the
rotation, regarding the collisional dynamics 
of hard spheres and a method for handling the inelastic collapse.  
The thermal reservoir of our system was 
modeled using white noise driving, first introduced by 
Williams and MacKintosh [18].  The strength of the noise was adjusted such 
that the average kinetic energy of all particle defines the kinetic 
temperature of the system.   

To test the theory properly, the system 
was started from a completely mixed state with no bias.  To accomplish 
this, the binary mixture was first heated to a very high temperature with no  
gravity in a closed  container.  This allowed the system to reach a proper 
steady state where the
centers of mass of both species A and B were the same, and the density profiles
of both species were uniform throughout the container.  Under these 
conditions, we achieved a perfectly mixed binary system.  We then turned on 
gravity, removed the lid of the container, and quenched
the system to a temperature, $T$, between the two critical
temperatures $T_c(A)$ and $T_c(B)$, such that $T_c(B)< T_c(A)$.
The system was then allowed to relax to
steady state under these new conditions.  Then we measured the
density profile, and monitored the center of mass.  The coefficient of 
restitution was 0.9999 for the particles and 0.98 for the walls, values 
where the system did not suffer from inelastic collapse [19].
We first present MD data for the usual Brazil nut problem [1].
Such a case is shown in Fig.\ 1
for $d_A/d_B = 8$, and $m_A/m_B = 4$ in 2D and $d_A/d_B = 2$, and $m_A/m_B = 
2$ in 3D.  
\begin{figure}[tb]
\begin{center}
\epsfig{file=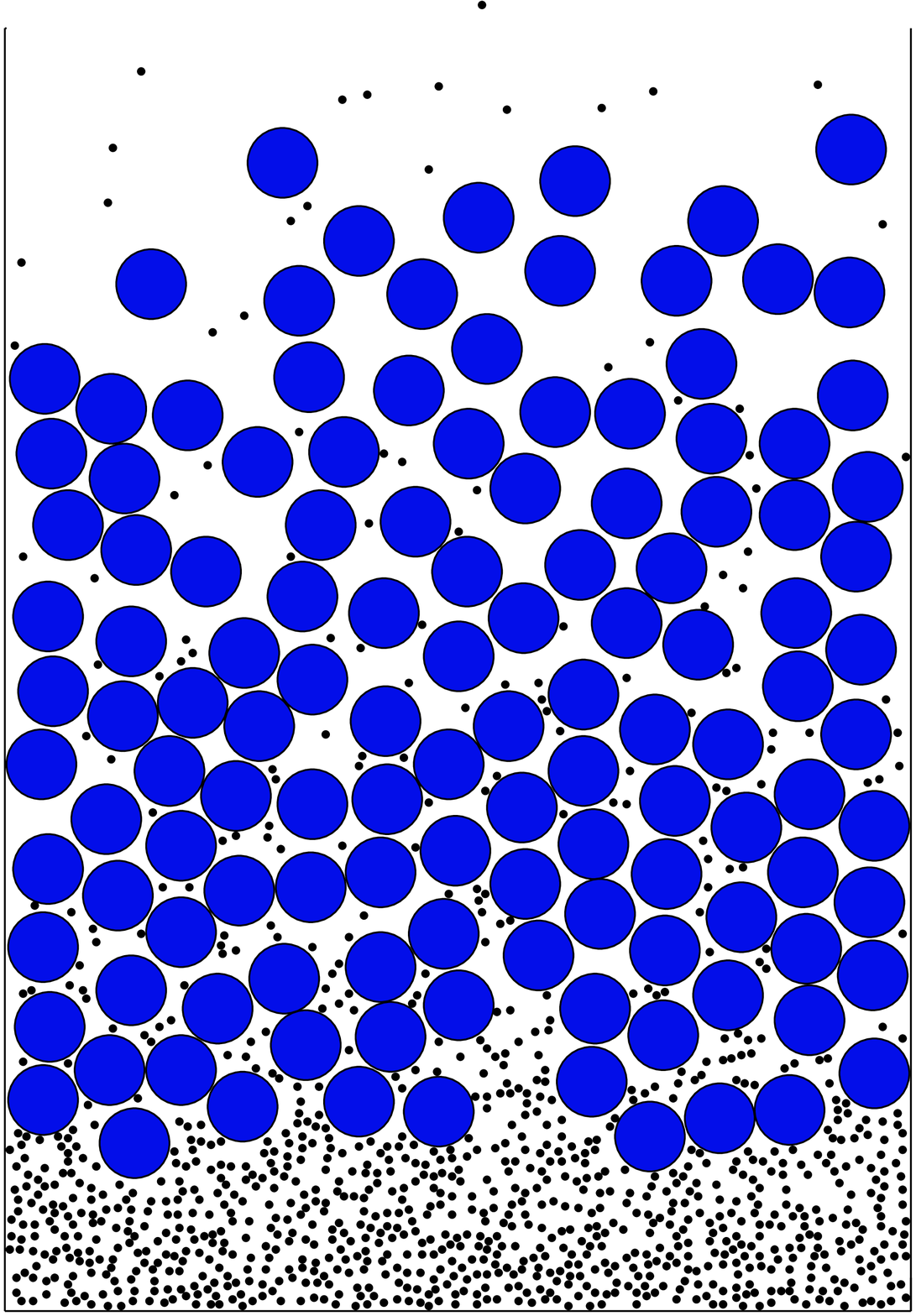,height=5cm,clip=}
\epsfig{file=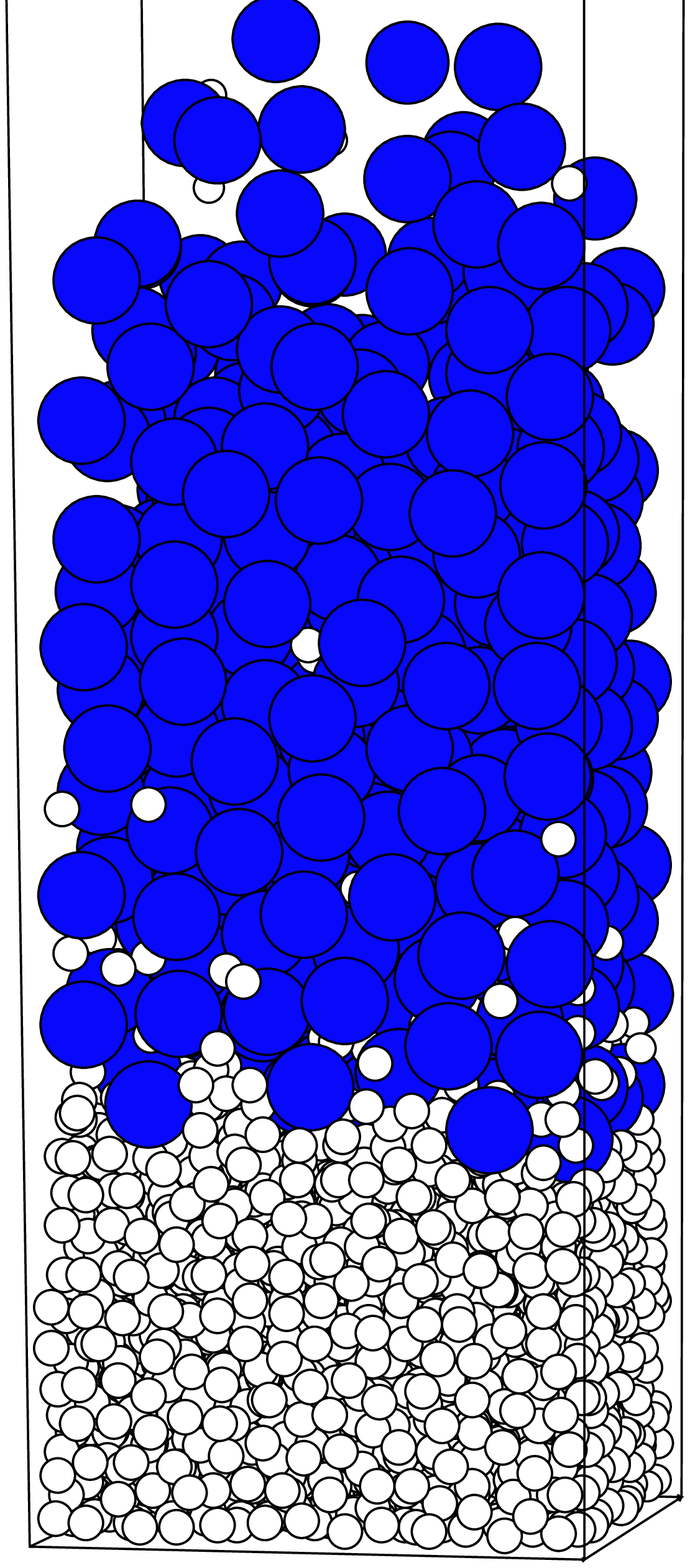,height=5cm,angle=-0.75,clip=}
\end{center}
\caption{Brazil nut problem in 2D (a) and 3D (b). Particles with the larger 
diameter rise to the top. Note that $T_c(B)< T < T_c(A)$ 
and percolation overrides 
the condensation effect.}
\end{figure}
The transition temperature ratio
is $T_c(A)/T_c(B) = m_Ad_A/m_Bd_B = 8$ in 2D and $4$ in 3D, 
and the quenching temperature $T$ is
such that $T_c(B) < T < T_c(A)$.  In these cases, 
we expect the larger particles 
to condense first, while the smaller particles remain fluidized.  
Indeed, the center of mass
position of particles of type A quickly decays as predicted by the
condensation picture, but as time passes, the smaller particles of type B 
evantually pass through the holes and settle to the bottom because of the
percolation effect. Percolation overrides the condensation mechanism, 
resulting in the usual BNP in two and three dimensions.  Snap shots of the 
equilibrium state, which is 
the typical Brazil nut segregation [1,5,7], are shown 
in Fig.\ 1.  At this point, one has two choices leading to 
the RBNP, the schematic picture of which is shown in Fig.\ 2.
The first option is to fix the mass ratio and change the
diameter ratio (Path 1).
There exists a critical diameter ratio below which the 
RBNP sets in.  This was also observed in Ref.\ [6].  The second case is to fix 
the diameter ratio and change the mass ratio.(Path 2)
Again, there exists a critical mass ratio beyond which the RBNP sets in.  
Both cases exhibit a crossover from the percolation to the condensation driven
segregation. 
\begin{figure}[tb]
\begin{center}
\epsfig{file=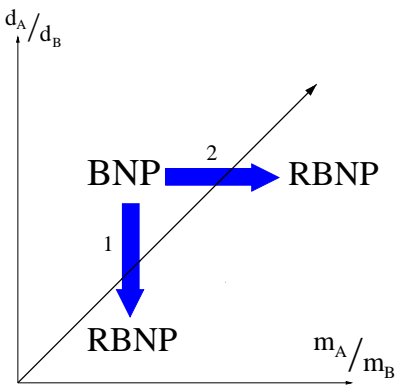,height=5cm,angle=0,clip=}
\end{center}
\caption{The schematic picture for the crossover from the Brazil
Nut Problem (BNP) to the Reverse Brazil Nut Problem (RBNP). The line
is the dividing line between the BNP and the RBNP for 2D.}
\end{figure}
We now present representative MD data to demonstrate this
crossover from BNP to RBNP.  We changed both the mass and the diameter, 
keeping $\mu_A=\mu_B$, to reverse the Brazil nut segregation via 
condensation.  In Fig.\ 3a, simulation results are shown for the same 
mass ratio as Fig.\ 1a, i.e $m_A/m_B=4$, but a diameter ratio of
$d_A/d_B=2$.  This represents movement along path 1 in the parameter 
space of Fig.\ 2. We quenched the system between the two critical temperatures 
such that $T_c(B) < T < T_c(A)$.  In this case, the larger particles condensed 
first and settled to the bottom.  They overcame the percolation effect and 
{\it remained} there, even though the diameter was larger.  In Fig.\ 3a, the 
simulation results clearly show the RBNP, namely,
the larger ones settling to the bottom.

One may also observe the RBNP by moving down along path 2
as
shown in Fig.\ 2.  The corresponding simulation result for the RBNP
is shown in Fig.\ 3b, where the diameter ratio is $d_A/d_B=2$,
but the mass ratio is $m_A/m_B=6$. 

The simulation results clearly suggest that during the segregation 
process, both the percolation effect and
the condensation mechanism compete.  Hence, the condition for
the crossover from the BNP to the RBNP may be determined
by setting the control parameters for both cases to be of 
the same order of magnitude.  

\begin{figure}[tb]
\begin{center}
\epsfig{figure=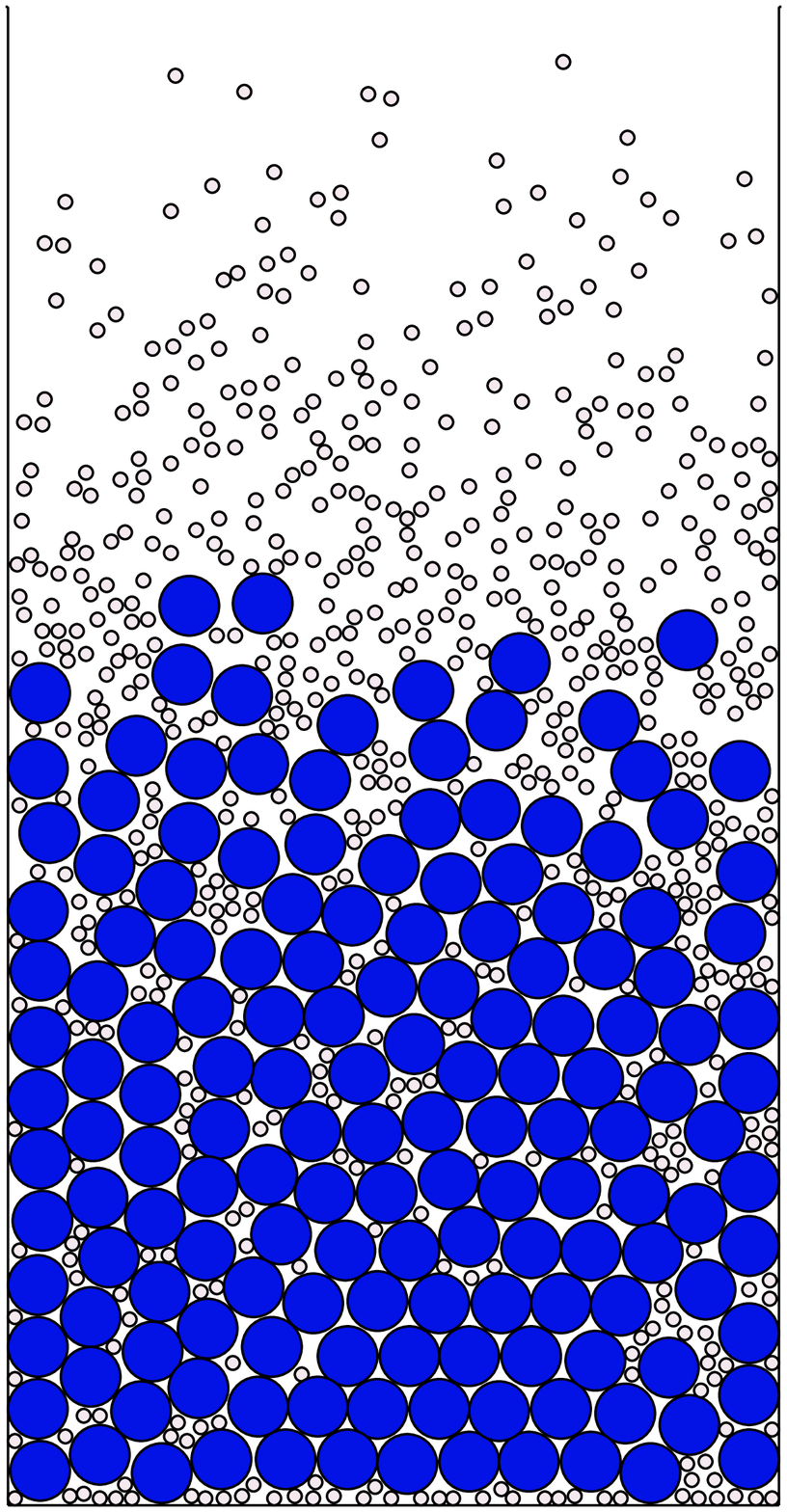,height=5cm,clip=}
\epsfig{figure=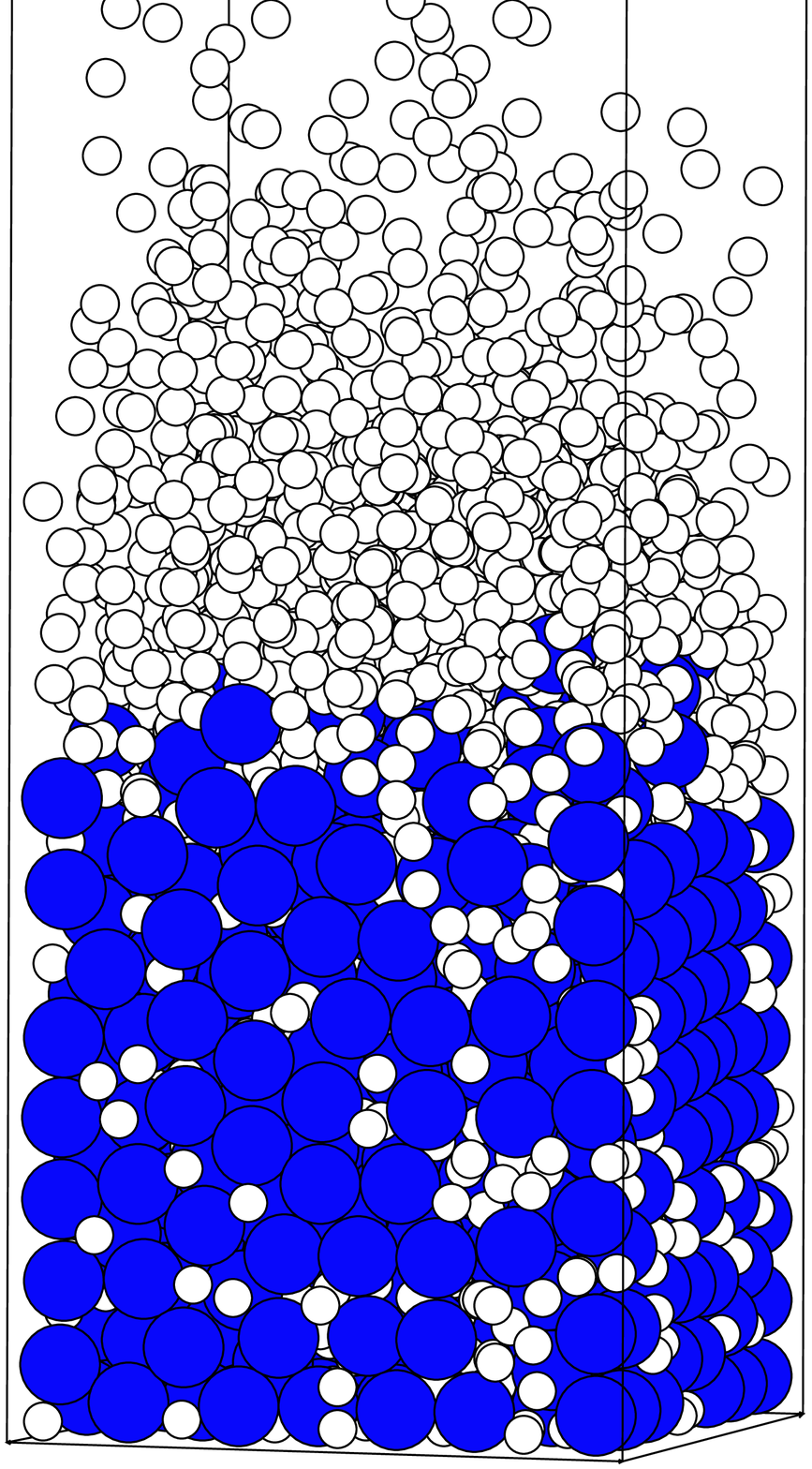,height=5cm,clip=}
\end{center}
\caption{(a) RBNP along path 1 in 2D.  Particles with the larger
diameter sink to the bottom. 
The mass ratio is the same as in Fig.\ 1a, but the
diameter ratio decreased to $d_A/d_B=2$. (b) RNBP
along path 2 in 3D.  The diameter ratio is the same as
in Fig.\ 1b, but the mass ratio is increased to $m_A/m_B=6$}.
\end{figure}

The control parameter for the percolation
effect has been determined by Rosato et al. [1], who argued that it is
basically controlled by the ratio of volumes (diameters) of the two types 
of particles.  Since we have established that the dimensionless
control parameter for condensation is
the ratio of the two critical temperatures, we determine
the crossover condition as follows:
$$ \left ( \frac{d_A}{d_B} \right )^D 
   \approx \frac{T_c(A)}{T_c(B)} = 
           \frac{m_Ad_A}{m_B d_B} ~,$$
or equivalently,
$$\left (\frac{d_A}{d_B} \right )^{D-1} \approx \frac{m_A}{m_B}~, \eqno (3)$$
where $D$ is the spatial dimension.  
The simulation results are displayed in Fig.\ 4a for two dimensions 
and Fig.\ 4b for three dimensions. 

The vertical axis is $y={d_A}/{d_B}$ and the
horizontal axis is $x={m_A}/{m_B}$.  Boundaries for the crossover
from the BNP to RBNP are given by the curve, $y=x^{1/(D-1)}$,
where $y=x$ in 2D and $y=\sqrt{x}$ in 3D.  
At the boundary line (stars), 
the system does not really segregate, but remains in a mixed state.
In passing, we point out that we
have also investigated the effect of the layer
thickness on segregation.  If we make $\mu_A>\mu_B$, such that
$T_c(B)<T<T_c(A)$, then we can {\it destroy} the BNP such that a mixed state
occurs.
\begin{figure}[tb]
\begin{center}
\epsfig{file=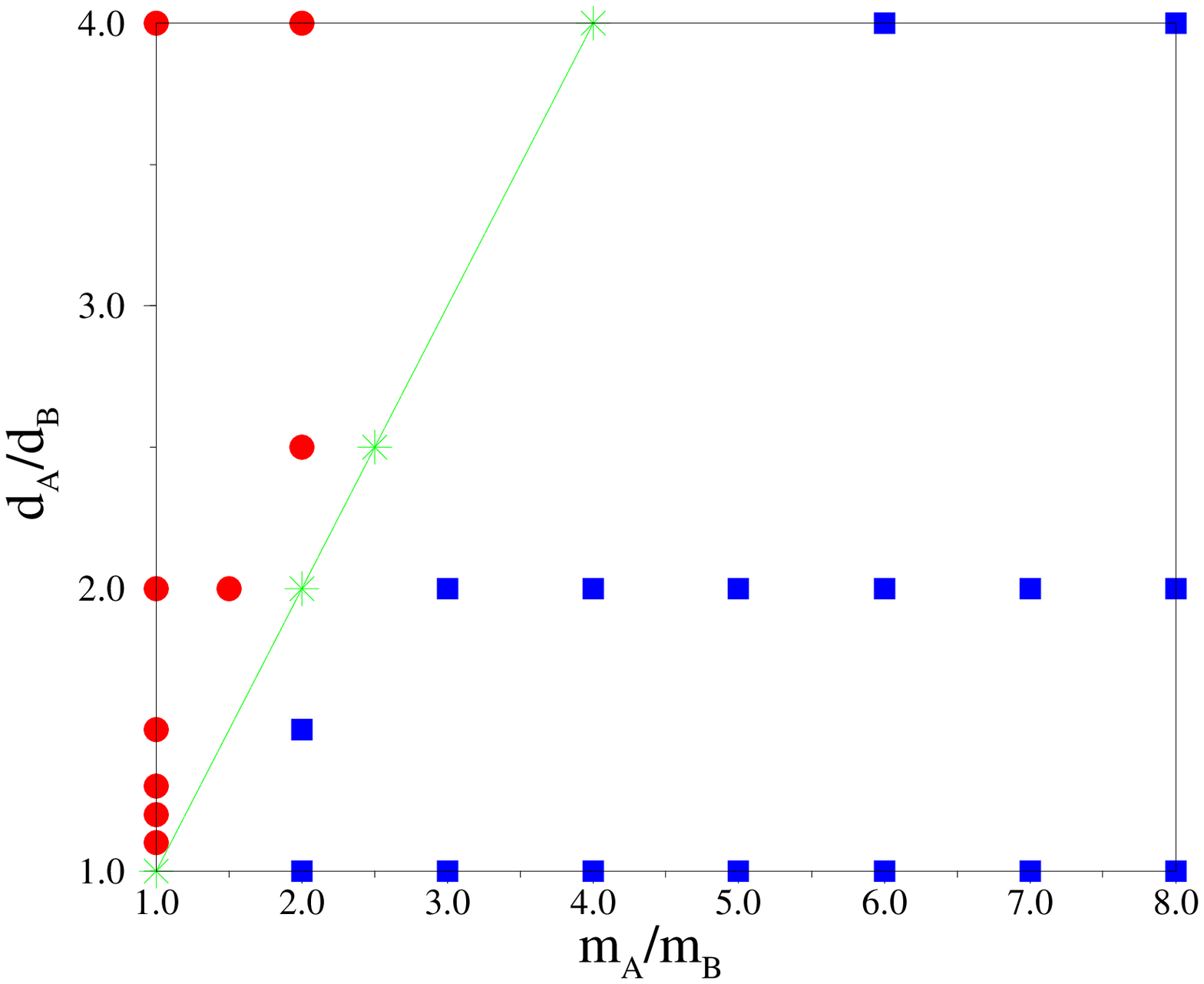,height=5 true cm,clip=}
\epsfig{file=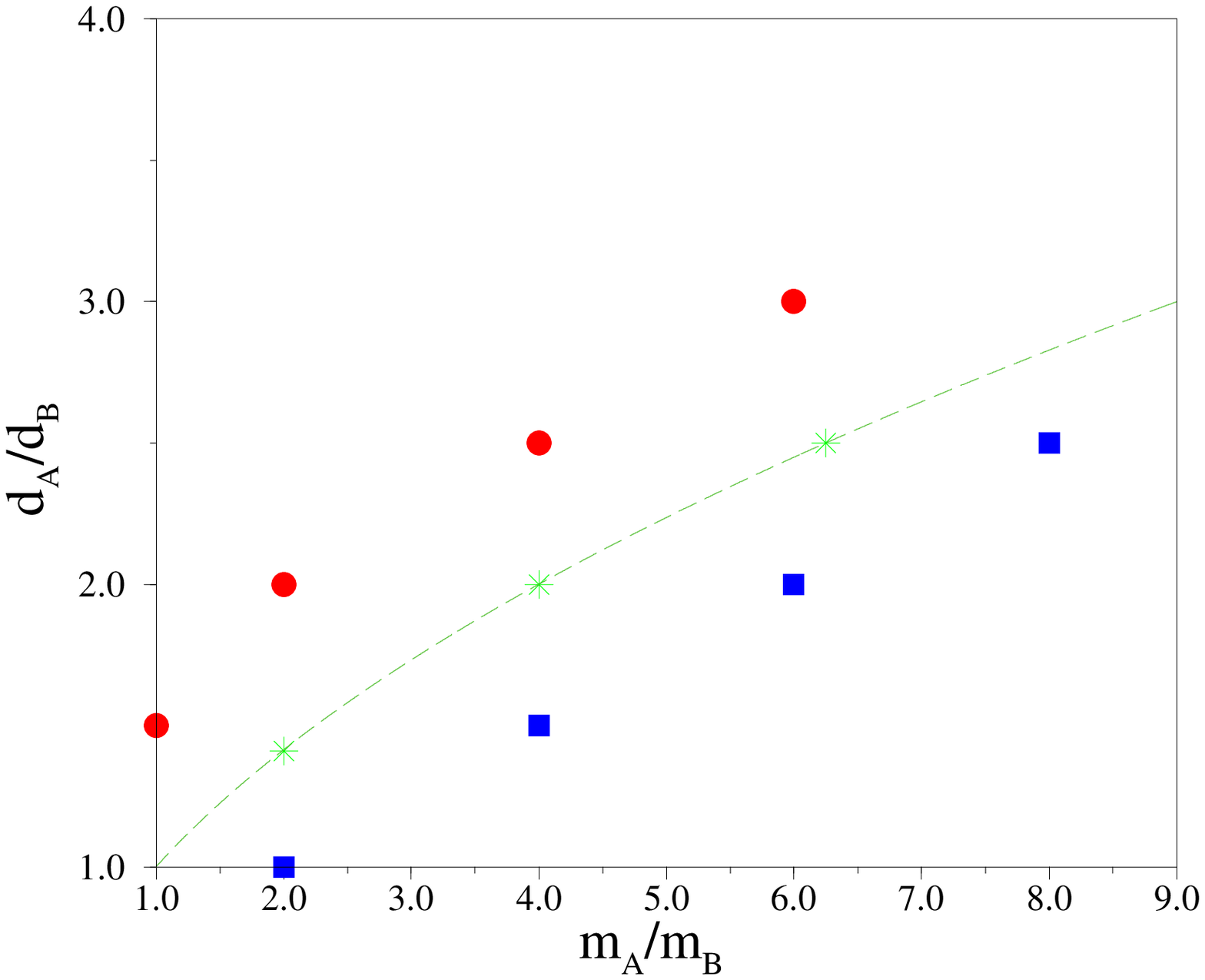,height=5 true cm,clip=}
\end{center}
\caption{Phase diagram determined for the segregation of binary mixtures
in (a) two dimensions, and (b) three dimensions.  In 2D, the straight line is
$y=x$, while in 3D, the curve is $y=\sqrt{x}$.  The circles are data points
for BNP, the squares for RBNP, and the stars are data points for 
the mixed state.}
\end{figure}

In summary, we have proposed a qualitative model for the segregation of 
binary mixtures and demonstrated that the type of segregation (BNP or RBNP) 
can be predicted.  The range of applications is beyond that of the classical 
percolation, reorganization, arching, and convection models since those
are limited to special situations.  For example, the arching model
requires very small temperatures or rather high densities in the 
condensed regime, whereas the convection model can only be applied
if convection cells are present.  However, we do not exclude those
approaches.  Rather, we expect that our condensation driven segregation
gives a more general framework, while special cases can be described better 
using the adapted models mentioned above.  We have related the segregation to
the respective condensation of two species, which takes place at different 
temperatures, and have studied various mass and size ratios confirming
the theoretical predictions via MD simulations in both two and three 
dimensions.  The future research needs to focus on
the effect of the filling heights on the condensation driven segregation
and examining
our strong assumption that the particles of either species feel
the other species only as phantom background.  

We wish to thank T. Shinbrot and Yan Levin for comments on the manuscript.
S. L. acknowledges the support of the Deutsche Forschungsgemeinschaft
(DFG).
\space
\newline

\noindent [1] T. Rosato et al., Phys. Rev. Lett. {\bf 58}, 1038 (1987).

\noindent [2] W. Reisner and E. Rothe, in {\it Bins and
Bunkers for Handling Bulk Materials}, series on Rock and Soil
Mechanics (Trans Tech, Clausthal-Zellerfeld, West Germany, 1971);
Z. T. Chowhan, Pharm. Technol. {\bf 19}, 56 (1995); S. Luding et 
al., Pharm. Technol. {\bf 20}, 42 (1996).

\noindent [3] J. C. Williams, Powder Technol. {\bf 15},
245 (1976); J. Bridgwater, Powder Technol. {\bf 15}, 215 (1976);
W. M. Visscher and M. Bolsterli, Nature(London) {\bf 239},
504 (1972).
 
\noindent [4] J. Knight, H. Jaeger and S. Nagel, Phys. Rev. Lett. {\bf 70},
3728 (1993). 

\noindent [5] J. Duran and J. Rajchenbach and E. Cl\'ement, 
Phys. Rev. Lett. {\bf 70}, 2431 (1993);  W. Cooke et al., Phys. Rev. E
{\bf 53}, 2812 (1996).

\noindent [6] R. Julien, P. Meakin, and A. Pavlovitvh, Phys. Rev. Lett.
{\bf 69}, 640 (1992).

\noindent [7] M. Dijikstra and D. Frenkel, Phys. Rev. Lett. {\bf 72},
298 (1994).

\noindent [8] Y-h. Taguchi, Phys. Rev. Lett. {\bf 69}, 1367 (1992);
See also, Int. J. Mod. Phys. A. {\bf 7}, 1839 (1993).

\noindent [9] D. C. Hong and S. Yue, Phys. Rev. E. {\bf 58}, 4763 (1998).

\noindent [10] Shinbrot and Muzzio, Phys. Rev. Lett. {\bf 81}, 4365 (1998).

\noindent [11] S. McNamara and S. Luding, in: {\it Segregation in Granular 
Flows}, ed. T. Rosato, IUTAM Symposium Proc. (Kluwer, Dordrecht, 2000); 
S. Luding, O. Strau\ss{} and S. McNamara, {\it ibid}.

\noindent [12] J. T. Willits and B. \"O. Arnarson, Phys. of Fluids
{\bf 11}, 3116 (1999).

\noindent [13] D. C. Hong, Physica A 271, 192 (1999).  

\noindent [14] P. V. Quinn and D. C. Hong, cond-mat/0005196.

\noindent [15] S. Luding and O. Strau\ss{}, in {\it Granular Gases},
T. P\"oschel and S. Luding, eds., Springer Verlag, Berlin (2000);
S. Luding, Europhys. Lett. (submitted).

\noindent [16] J. P. Hansen and I. R. McDonald, Theory of simple liquids,
Academic Press Limited, London (1986).

\noindent [17] B. D. Lubachevsky, J. Comp. Phys. 94, 255(1991);
S. Luding, Phys. Rev. E {\bf 52}, 4442 (1995);
S. Luding and S. McNamara, Granular Matter {\bf 1}(3), 113 (1998).

\noindent [18] D. M. Williams and F. C. MacKintosh, Phys. Rev. E 
{\bf 54}, R9 (1996).

\noindent [19] I. Goldhirsch and G. Zanetti, Phys. Rev. Lett. {\bf 70}, 
1619 (1993); S. McNamara and W. R. Young, Phys. Rev. E {\bf 53}, 5089 (1996). 
\end{document}